\documentclass[usenatbib,usegraphicx,useAMS]{mn2e}
\usepackage{times}

\def\vlsr {{\rm V}$_{\rm LSR}$}
\def\kms  {{\rm km~s}$^{-1}$}

\begin{document}

% 21sep04 draft
% 29sep04 draft

\title[Broad 6-cm OH lines]{VLA observations of broad 6-cm excited state OH lines in W49A}
\author[Palmer \& Goss] 
{Patrick Palmer,$^1$\thanks{E-mail: ppalmer@oskar.uchicago.edu},
W. M. Goss,$^2$\thanks{E-mail: mgoss@nrao.edu} \\
$^1$Department of Astronomy and Astrophysics, University of
Chicago, 5640 S. Ellis Ave., Chicago, IL 60637 \\
$^2$National Radio Astronomy Observatory, P. O. Box O, Socorro, NM 87801 \\ 
}

\maketitle
\begin{abstract}
Using the Very Large Array (VLA), we observed all three of the 6-cm lines
of the $^2\Pi_{1/2},J=1/2$ state of OH with sub-arcsecond resolution
($\sim$0.4 arcsec) in W49A.  While the spatial distribution and the range
in velocities of the 6-cm lines are similar to those of the ground state
(18-cm) OH lines, a large fraction of the total emission in all three 6-cm
lines has large linewidths ($\sim$5 -- 10 \kms) and is spatially-extended,
very unlike typical ground state OH masers which typically are point-like
at VLA resolutions and have linewidths $\leq$1 \kms.  We find brightness
temperatures of 5900 K, 4700 K, and $\geq$730 K for the 4660-MHz, 4750-MHz,
and 4765-MHz lines, respectively.  We conclude that these are indeed maser
lines.  However, the gains are $\sim$0.3, again very unlike the 18-cm lines
which have gains $\geq$10$^4$.  We compare the excited state OH emission
with that from other molecules observed with comparable angular resolution
to estimate physical conditions in the regions emitting the peculiar,
low-gain maser lines.  We also comment on the relationship with the 18-cm
masers.
\end{abstract}

\begin{keywords}  
masers -- ISM:molecules -- ISM:individual(W49A) -- radio lines:ISM
\end{keywords}

\section{INTRODUCTION}
The $^2\Pi_{1/2},J=1/2$ rotationally excited state of OH has three
hyperfine transitions in the 6-cm band: F=1$\rightarrow$0 ($\nu_0$=
4765.562 MHz), F=1$\rightarrow$1 ($\nu_0$=4750.656 MHz), and
F=0$\rightarrow$1 ($\nu_0$=4660.242 MHz).  Emission from this state of OH
was discovered \citep{zppl68} and is most frequently observed
(\citealt*{cmw91,cmc95}; \citealt{de02,s03}) as maser emission in the
4765-MHz line.  The linewidths are typically $<$1 km s$^{-1}$.  Another
type of emission was detected with linewidths of a few to 20 km s$^{-1}$,
first in the 4660-MHz line in Sgr B2 \citep*{grs71} and shortly afterward
in both the 4750-MHz and 4765-MHz lines as well \citep{gr71}.  The broad
lines had much lower peak flux densities than the narrow 4765-MHz lines
which are typically observed to have peak densities exceeding 1 Jy.
Subsequently, \citep*{rzp75,rpz80,gm83} observed emission with large line
widths in a number of sources including W49A, the subject of this paper.
The differences between the two types of emission seemed clearcut: 1) large
linewidths vs. narrow linewidths; 2) angular sizes measurable with single
dish telescopes vs. point-like sources; and 3) low peak luminosities versus
high.

After the detection of analogous broad emission lines from the
$^2\Pi_{3/2},J=5/2$ near 5-cm wavelength in Sgr B2, as well as broad
absorption features in the 5-cm lines in W3(C), \citet{rzp75} suggested
that the broad lines (emission and absorption) be called ``quasi-thermal'':
i.e. close to thermal excitation (which included low gain masers) in
contrast to the excitation of strong, high gain masers such as the ground
state 18-cm lines, the narrow 4765-MHz lines, and the intense,
narrow 5-cm and 2-cm OH masers \citep*{yzpp69,tpz70}.

The suggestion of \citet{gm83} that both masers and quasi-thermal emission
originate in the same clouds, i.e. ``that masers are hot spots within an
envelope defined by the extent of the quasi-thermal emission'' could only
be checked by interferometric observations.  \citet*{gbw85} carried out
observations of 4750-MHz emission and 4660-MHz absorption with the VLA
which, when combined with 4765-MHz observations of \citet*{gwp83}, showed that
this view was plausible for W3(OH).  Subsequent interferometric studies
(Sgr B2: \citealt*{gwp87}; W49A: \citealt*{pgw04}, hereafter: PGW), showed that
while the broad 4660- and 4750-MHz emission regions were smaller than
estimated from single dish studies, they were spatially resolved, unlike
the narrow linewidth 4765-MHz masers, many of which have angular sizes
$\sim$5 mas (\citealt*{pgd03}, hereafter PGD).

Theoretical models to explain emission and absorption in a number of
excited state OH lines simultaneously for several sources have included
all lines in the analysis \citep{cw91,jfgw94}.  However,
the 10-20 km s$^{-1}$ wide lines observed in W49A and Sgr B2 have not yet
been successfully modelled.

In this study we have observed all three 6-cm excited state OH lines from
W49A with the VLA to determine if the broad line
sources would break up into small clumps of narrower linewidth features
when observed with resolution $\leq$1 arcsec.  In addition, we planned to
determine the brightness temperatures and to determine precisely, both
spatially and in velocity, the relationship between emission in these
OH lines and emission and absorption in other molecular lines and
the ionized gas.  Finally, we address the origins of these
broad lines.

\section{OBSERVATIONS}

The observations were conducted in four sessions using the VLA of the
National Radio Astronomy Observatory\footnote{The National Radio Astronomy
Observatory is a facility of the National Science Foundation operated under
cooperative agreement by Associated Universities, Inc.}  .  The bulk of the
data reported here are the high angular resolution A- and BnA-configuration
observations obtained in 2002.  Brief observations with lower angular
resolution were conducted in 2001.  The pointing positions and centre
\vlsr's are provided in Table~\ref{tab:tab1}; other parameters of these
observations are summarized in Table~\ref{tab:tab2}.  The distance to W49A
is 11.4 kpc \citep*{gmr92}.

On 2002 April 1, the 4660- and 4750-MHz lines were observed for
alternating 20 minute intervals, separated by a 3 minute observation of the
phase calibrator at each frequency.  This scheme allowed coverage of an
hour angle range of about $\pm$3.5 hours for each line.  The bandpass
calibrator was observed for 11 minutes at each frequency at the beginning
of the session.  The 4765-MHz line was observed on 2002 June 1 following an
analogous strategy: a 17 minute observation on source was followed by a 3
minute observation of the phase calibrator.  However, the hour angle
coverage was less extensive (+2.1 hour to +4.4 hours) so that imaging
quality is not as favourable for the continuum sources, but is
noise-limited for the compact maser regions.  The bandpass calibrator was
observed for 13 minutes at the end of the session.  Both 2001 observations
consisted of two scans separated by about an hour and took place at rather
large hour angles.  The bandpass calibrator used on each date is listed in
Table~\ref{tab:tab2}.  The same phase calibrator was used in all
observations (1821+107).  The flux density scale was set by observations of
3C286 (April 1 and August 27) or 3C48 (June 1 and November 15).

The correlator setup generated spectra with 127 independent Hanning
smoothed channels with widths 12.207 kHz.  The velocity resolution differed
slightly among the three lines because of the difference in rest
frequencies.  The velocity resolution is shown in the fourth column of
Table~\ref{tab:tab2}.  The total velocity range covered in these
observations was $\sim \pm$40 \kms.

All data were reduced with the NRAO's AIPS reduction package.  For the
large datasets obtained in 2002, two sets of images were made.  One set
was made at the full resolution available in the dataset (corresponding to
the synthesised beams provided in the last column of Table~\ref{tab:tab2}).
In addition, the 4660- and 4750-MHz images (2002 April 1) were convolved to
1-arcsec  resolution to improve the signal-to-noise for low brightness
emission; the 4765-MHz image (2002 June 1) was convolved to 2-arcsec 
resolution both to improve the signal-to-noise and to produce a circular beam
(in the BnA data, the beam was elongated by about 3:1).  The full-resolution
images were primarily used to determine properties of small sources; the
convolved images were primarily used to study the relatively low brightness
extended OH emission lines.

\section{RESULTS}

All three of the 6-cm OH lines in W49A are found to have large linewidth,
spatially extended components and narrow linewidth, small or point-like
components.  Fig.~\ref{fig:fig1}(a) is an image of the 3.6-cm continuum
emission reported by \cite*{dmg97} (hereafter DMG).  The components
discussed in this paper are indicated by a cross and labelled (using the
names provided by DMG).  The 6-cm OH line sources are observed at a number
of positions in central 2 -- 3 arcmin in W49A, but the large
linewidth, spatially-extended components are observed only in a region
about 15 x 6 arcsec$^2$ extending E-W over sources G, E, B, and A of
DMG.  This region is the primary focus of this paper and we will refer to
it as the ``extended OH region''.  Furthermore, even within this small
region, most of the line flux density originates from the $\sim$7 
x $\sim$5 arcsec$^2$ region of source G.  A number of small angular size
sources were observed in all three lines both within and outside the
extended OH region.  The extended OH region will be discussed in Section
3.1; the 6-cm OH emission sources outside of this region will be discussed
briefly in Section 3.2.

\subsection{Extended OH region}
The bulk of the emission in all three lines is confined to the source G
region.  Fig.~\ref{fig:fig1}(b) -- (d) show the zeroth moment of the 4660-,
4750-, and 4765-MHz OH emission superimposed on the continuum.  Both
extended and point-like components are observed.  While a large fraction of
the emission arises from the source G region, emission continues westward
across source B in all three lines; in the 4660- and 4765-MHz lines
emission continues farther westward across source A.  Roughly, the line
emission in these regions follows the continuum emission.  However, in
detail there are numerous differences among the three line images and
between each of the line images and the continuum.  For example, the
4750-MHz image shows a secondary maximum near source B which is not evident
in either the 4660- or the 4765-MHz data.  In the full resolution data,
while all three lines show a number of point-like sources, the current high
spatial resolution data shows that the emission cannot be accounted for by
a collection of point-like sources.  A significant fraction of the emission
in all three lines is spatially extended (see also Section 4.2).

In order to analyse the extended emission at 4660- and 4750-MHz, we used
the AIPS program XGAUS to fit single Gaussians at each pixel (RA, Dec) for
which three consecutive velocity channels had signals $\geq$3 $\sigma$.
Therefore, XGAUS discriminates against very narrow lines; such lines were
located by visual inspection.  The fits were done on images convolved to
1-arcsec resolution and the results are displayed in
Fig.~\ref{fig:fig3}.  For each line, the amplitude, \vlsr, and FWHM at
each pixel are plotted.  (In a few small areas there are clearly two
components; these were fit separately, but are not displayed.)

\subsubsection{Comparison with earlier results}

Inspection of Fig.~\ref{fig:fig1}(a) -- (c) shows that the angular sizes
$\sim$30 arcsec reported in early single dish studies were a result of beam
smearing.  Even the smallest single dish beam used [2.6 arcmin
(\citealt{rpz80}, hereafter: RPZ)] blended together sources A -- G; and, at
4660- and 4765-MHz, the 2.6-arcmin beam blended in other OH line
emission in the field.

To facilitate comparisons with earlier single dish results, the line
profiles integrated over the extended OH region (sources G -- A) are
presented in Fig.~\ref{fig:fig2}.  Fig.~\ref{fig:fig2}(a) -- (c) show the
integrated profiles from the high spatial resolution 2002 data, and
Fig.~\ref{fig:fig2}(d) and (e) display the integrated profiles from the
lower spatial resolution 2001 data.  In these spatially averaged spectra,
only the 4765-MHz spectrum [Fig.~\ref{fig:fig2}(c)] contains obvious
narrow components at \vlsr $\sim$2.3 and $\sim$16.0 \kms.  The
spectral baseline has a rather large slope in Fig.~\ref{fig:fig2}(a) and
(b).  This slope is due to the missing short spacings in the A-array.  The
spectral baseline slope in Fig.~\ref{fig:fig2}(e) is present because no
bandpass calibrator was observed on that date.

The peak flux densities in all three lines from earlier measurements and
from this paper are provided in Table~\ref{tab:tab3}.  Values quoted
without error bars were read from plots with poorly determined precision.
The VLA spectra were integrated over a box which includes sources G -- A
and is therefore larger than the synthesised beam.  For the VLA
observations, the size of the box used is tabulated in the fourth column of
Table~\ref{tab:tab3}.  In the D-array data (2001 Nov 15) a noticeably
larger box was used because the synthesised beam was $\sim$16 arcsec.  The
single dish spectra differ in character from the VLA spectra because the
single dish beams include varying contributions from sources outside of the
extended OH region. Because the VLA spectra are constructed by integrations
over many synthesised beams, they are unaffected by emission outside of the
extended OH region.  This difference is not significant for extended
component measurements except at 4765-MHz.  Inspection of
Table~\ref{tab:tab3} shows that there is no evidence for variability of the
broad components --- very unlike the behaviour of the narrow components at
4765-MHz (see PGW).  In addition, because the flux densities are so similar
for the entire range of resolutions, there is no evidence that a
significant amount of the flux density is resolved by the A-array (i.e no
evidence that there is a component of OH emission with size $\geq$10 arcsec).

\subsubsection{4660-MHz}

Inspection of Fig.~\ref{fig:fig3}(a) reveals OH emission at or near the
positions of DMG sources A, B, E, and throughout the G complex.  Within
this complex, the most intense OH lines arise near G$_2$.
Fig.~\ref{fig:fig3}(c) and (e) show that there is a large velocity gradient
across source A and that the FWHM is largest ($\geq$15 \kms) in the western
part of this source. Source B has the most extreme velocity in the region
with a smaller FWHM ($\sim$4 \kms).  Source E has an intermediate \vlsr \
and FWHM ($\sim$7 \kms).  Finally, within source G, \vlsr \ and FWHM vary by
significant amounts, but lie within the range of values observed in the
other sources.

In a spectrum integrated over the source A region at 4660-MHz, two
components are separated in velocity by $\sim$10 \kms [see
Fig.~\ref{fig:fig4}(a)].  The presence of two components causes the
apparently extreme FWHM for this component in Fig.~\ref{fig:fig3}(e).
However, fits on a pixel-by-pixel basis in the full resolution data reveal
the apparent duplicity to be due to two spatially unresolved components
separated by $\sim$0.15 arcsec.  The position, peak flux density, \vlsr,
and FWHM of these components are summarised in Table~\ref{tab:tab4}.

The spectrum integrated over the source B region is shown in
Fig.~\ref{fig:fig4}(b). At some velocities there are two components,
separated by $\sim$0.8 arcsec.  The velocities differ by $\sim$0.2
\kms.  However, unlike source A, both components are spatially resolved
with sizes $\sim$0.5 arcsec.  However, because of low signal-to-noise, the
two components are treated as a single entity in further fits.  The
position, tabulated in Table~\ref{tab:tab4}, and an angular size
$\sim$1 arcsec were determined from a Gaussian fit to the relevant portion
of Fig.~\ref{fig:fig3}(a).  The velocity and FWHM in Table~\ref{tab:tab4}
are averages of the central pixels of Fig.~\ref{fig:fig3}(c) and (e).  The
spectrum integrated over source E is shown in Fig.~\ref{fig:fig4}(c).
Because its peak flux density is only $\sim$10 mJy and because it is not
well resolved from sources B and G, parameters are not tabulated.

The \vlsr \ and FWHM in source G are displayed in Fig.~\ref{fig:fig5}(a)
and (b).  Both quantities vary across the source.  The linewidth ranges
from $\sim$2
\kms \ to $\sim$20 \kms.  \vlsr \ ranges from $\sim$2 \kms \ to $\sim$8
\kms.  Fig.~\ref{fig:fig6}(a) shows the spectrum integrated over the whole
of source G; Fig.~\ref{fig:fig6}(b) -- (f) are the spectra integrated over
small regions (0.75 arcsec square) at the positions of sources G$_1$ --
G$_5$.  Note that all of these spectra are well resolved in velocity with
FWHM's ranging from $\sim$6 \kms \ to $\sim$15 \kms.  Because we will
compare the OH emission from the source G region with emission and
absorption from other molecules, we tabulate the peak flux density, \vlsr,
and FWHM in Table~\ref{tab:tab5} below.  These values were determined from
the profile shown Fig.~\ref{fig:fig6}(a).

\subsubsection{4750-MHz}

Fig.~\ref{fig:fig3}(b) shows that the 4750-MHz emission has maxima at sources
B, E, and at several points in the G region.  Source A is not detected at
4750-MHz.  The 4660-MHz and 4750-MHz velocity distributions
[Fig.~\ref{fig:fig3}(c) and (d)] are similar.  However, in a band running
between sources G$_2$ and G$_3$ through source G$_4$, the 4750-MHz line is
noticeably broader with FWHM as large as $\sim$18 \kms \ [compare
Fig.~\ref{fig:fig3}(e) and (f)].

The spectrum integrated over the source B region is shown in
Fig.~\ref{fig:fig4}(d).  Due to less favourable signal-to-noise at 4750-MHz, 
it is not possible to separate spatial components as was done for 4660-MHz.
The position listed in Table~\ref{tab:tab4} was derived from a Gaussian fit
to the source B region of Fig.~\ref{fig:fig3}(b).  The velocity and FWHM
are averages of the central pixels of Fig.~\ref{fig:fig3}(d) and (f).  Unlike
the 4660-MHz emission from this source, there is no evidence for double
lines, and no significant velocity gradient is observed.

At 4750-MHz, emission from source E is more intense and more spatially
distinct from B and G than at 4660-MHz.  The spectrum integrated over
source E is shown in Fig.~\ref{fig:fig4}(e).  Line parameters and
positional information are derived as for source B and tabulated in
Table~\ref{tab:tab4}.

As at 4660-MHz, the 4750-MHz emission in the source G region is quite
complex.  The \vlsr \ and FWHM are displayed in Fig.~\ref{fig:fig5}(c) and
(d).  The ranges of \vlsr \ and FWHM are similar to the ranges observed at
4660-MHz.  Fig.~\ref{fig:fig7}(a) shows a spectrum integrated over source
G; Fig.~\ref{fig:fig7}(b) -- (f) show spectra in the directions of the five
DMG components (integrated over boxes at the same positions and having the
same sizes as used at 4660-MHz).  Compared with similar profiles at
4660-MHz (Fig.~\ref{fig:fig6}), small but significant differences are
observed.  As at 4660-MHz in this region, all of these spectra are resolved
in velocity with FWHM's ranging from $\sim$5 \kms \ to $\sim$17 \kms.
Parameters of the line profile integrated over source G
[Fig.~\ref{fig:fig7}(a)] are presented in Table~\ref{tab:tab5}.

\subsubsection{4765-MHz}

\begin{figure*}
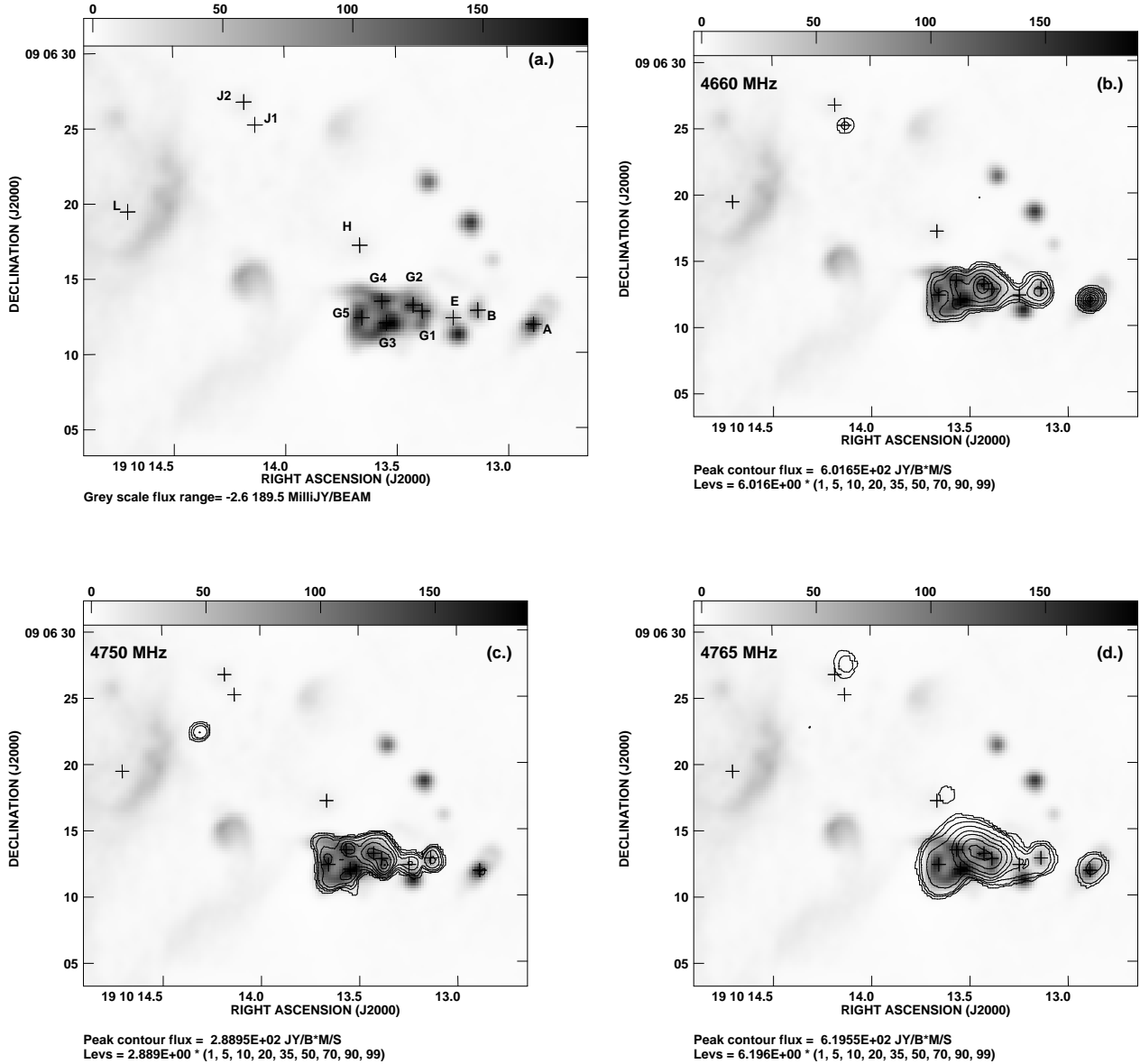

\centering
\begin{minipage}[c]{.45\textwidth}
   \centering
   \includegraphics[width=\textwidth,angle=-90]{fig1a.eps}
\end{minipage}
\hspace{0.05\textwidth}
\begin{minipage}[c]{.45\textwidth}
   \centering
   \includegraphics[width=\textwidth,angle=-90]{fig1b.eps}
\end{minipage}\\[4truemm]
\begin{minipage}[c]{.45\textwidth}
   \centering
   \includegraphics[width=\textwidth,angle=-90]{fig1c.eps}
\end{minipage}
\hspace{0.05\textwidth}
\begin{minipage}[c]{.45\textwidth}
   \centering
   \includegraphics[width=\textwidth,angle=-90]{fig1d.eps}
\end{minipage}
\caption{Comparison of 3.6-cm continuum emission from DMG with the
velocity-integrated OH emission at 4660-, 4750-, and 4765-MHz in the
extended OH region of W49A.  (a) The 3.6-cm continuum emission from DMG is
shown in grey scale. For the sources discussed in the text, the position
for the continuum components provided by DMG is indicated by a cross and
labelled.  (L is a diffuse source of which only a bright arc is evident in
this high resolution data.  The position shown was determined from low
resolution data and refers to the whole source.)  The synthesised beam for
this image is 0.8 arcsec.  (b) The zeroth moment of the 4660-MHz OH
emission on 2002 April 1 is represented by contours superimposed on the
gray scale of the continuum.  (c) The zeroth moment of the 4750-MHz OH
emission on 2002 April 1 superimposed as in (b).  In both (b) and (c), the
OH image has been convolved to a circular beam with FWHM= 1.0 arcsec.  (d)
The zeroth moment of the 4765-MHz emission on 2002 June 1 represented by
contours is superimposed on the gray scale of the continuum.  The OH image
has been convolved to a circular beam with FWHM= 2.0 arcsec.  The contour
levels are 1, 5 10, 20, 35, 50, 70, 90, and 99 per cent of the maximum in
each OH image.}
\label{fig:fig1}
\end{figure*}

\begin{figure*}
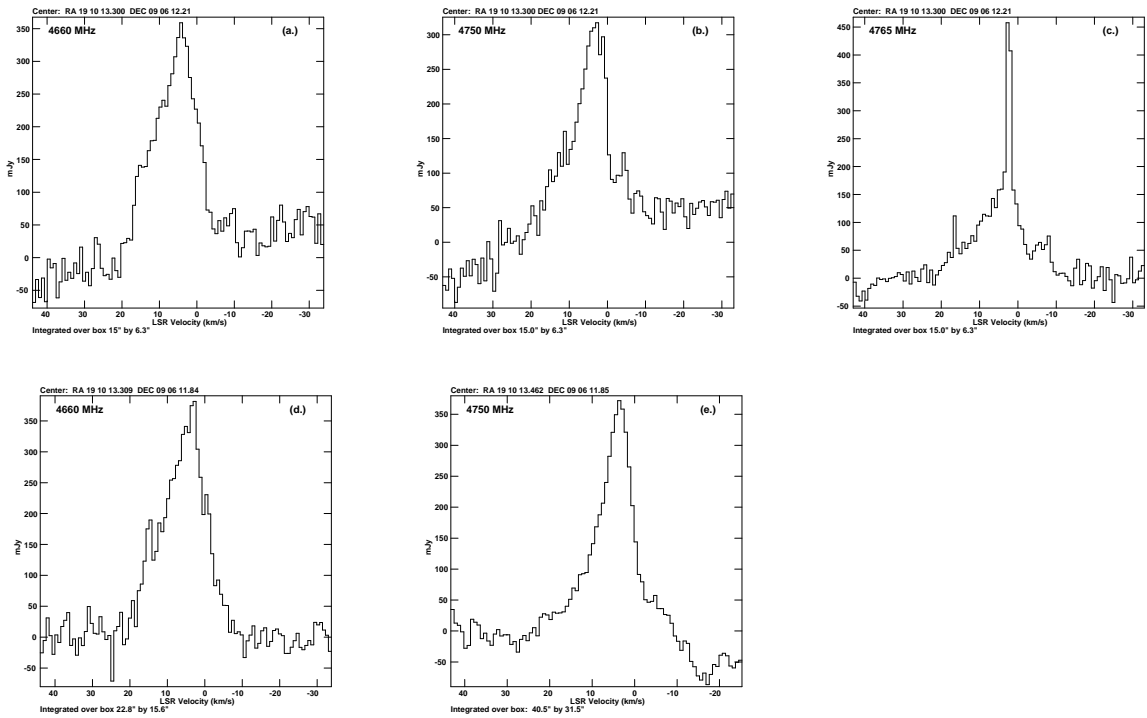

%\centering
\begin{minipage}[c]{.25\textwidth}
   \centering
   \includegraphics[width=\textwidth]{fig2a.eps}
\end{minipage}
\hspace{0.05\textwidth}
\begin{minipage}[c]{.25\textwidth}
   \centering
   \includegraphics[width=\textwidth]{fig2b.eps}
\end{minipage}
\hspace{0.05\textwidth}
\begin{minipage}[c]{.25\textwidth}
   \centering
   \includegraphics[width=\textwidth]{fig2c.eps}
\end{minipage}\\[4truemm]
\hspace{-.30\textwidth}
\begin{minipage}[c]{.25\textwidth}
%   \centering
   \includegraphics[width=\textwidth]{fig2d.eps}
\end{minipage}
\hspace{0.05\textwidth}
\begin{minipage}[c]{.25\textwidth}
%   \centering
   \includegraphics[width=\textwidth]{fig2e.eps}
\end{minipage}
\caption{The 6-cm OH spectra integrated over regions covering sources G -- A 
in the extended OH region.  The dimensions of the regions are provided in
Table~\ref{tab:tab3}.  (a) The 4660-MHz line observed in 2002; (b) the
4750-MHz line observed in 2002; (c) the 4765-MHz line observed in 2002
(with narrow components near \vlsr=2.3 \kms \ and \vlsr=16 \kms); (d) the
4660-MHz line observed in 2001; and (e) the 4750-MHz line observed in 2001.
The baseline is truncated in (e) because of severe curvature at the low
velocity end due to the lack of bandpass calibration at this date.}

\label{fig:fig2}
\end{figure*}

\begin{figure*}
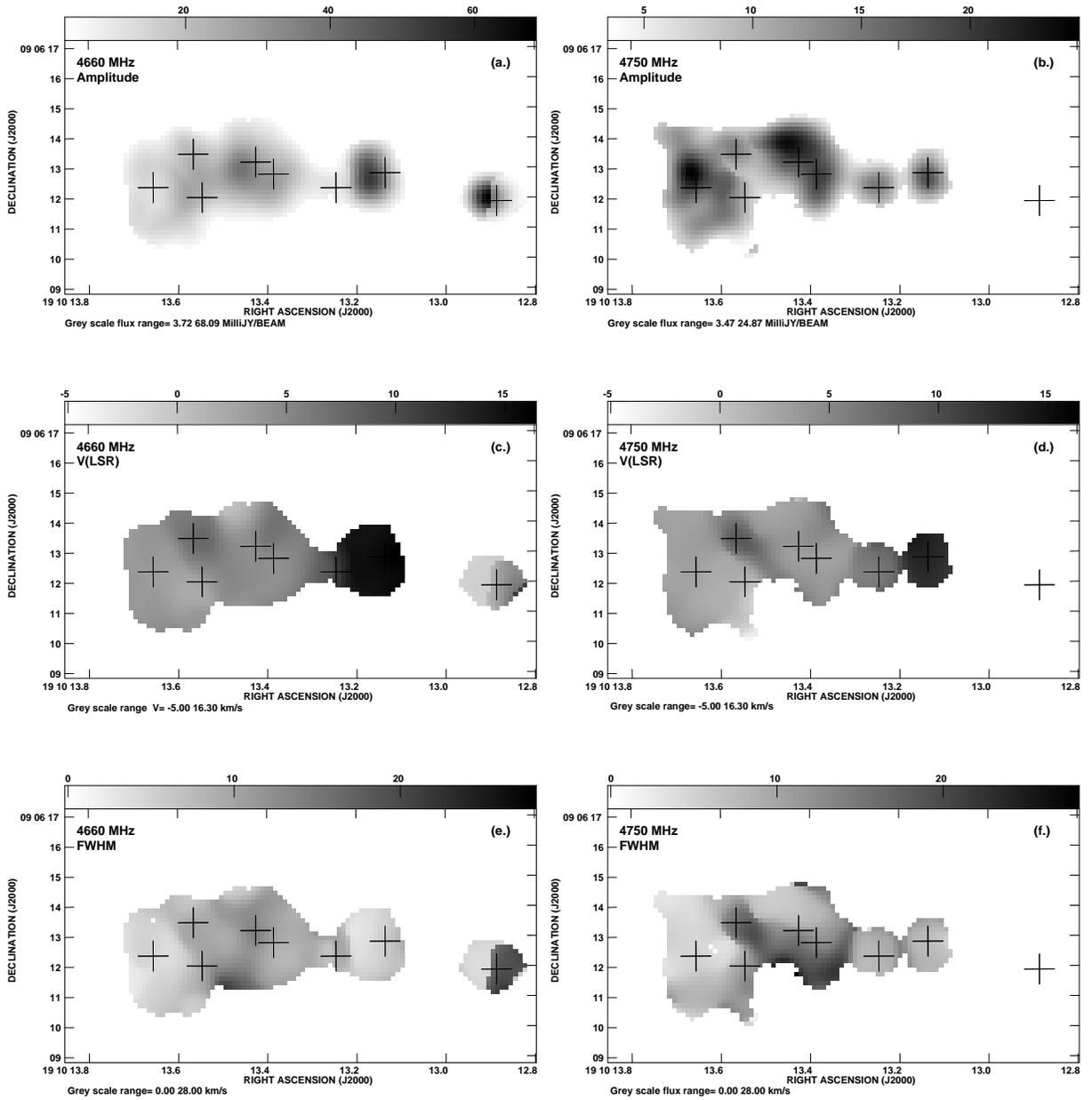

\centering
\begin{minipage}[c]{.3\textwidth}
   \centering
   \includegraphics[width=\textwidth,angle=-90]{fig3a.eps}
\end{minipage}
\hspace{0.15\textwidth}
\begin{minipage}[c]{.3\textwidth}
   \centering
   \includegraphics[width=\textwidth,angle=-90]{fig3b.eps}
\end{minipage}\\[4truemm]
\begin{minipage}[c]{.3\textwidth}
   \centering
   \includegraphics[width=\textwidth,angle=-90]{fig3c.eps}
\end{minipage}
\hspace{0.15\textwidth}
\begin{minipage}[c]{.3\textwidth}
   \centering
   \includegraphics[width=\textwidth,angle=-90]{fig3d.eps}
\end{minipage}\\[4truemm]
\begin{minipage}[c]{.3\textwidth}
   \centering
   \includegraphics[width=\textwidth,angle=-90]{fig3e.eps}
\end{minipage}
\hspace{0.15\textwidth}
\begin{minipage}[c]{.3\textwidth}
   \centering
   \includegraphics[width=\textwidth,angle=-90]{fig3f.eps}
\end{minipage}
\caption{Images created from the results of Gaussian fits at each pixel in the
source G -- A region. The images were convolved to 1 arcsec resolution
before fitting.  Displayed are: the peak amplitude at 4660-MHz (a) and at
4750-MHz (b); the LSR velocity at 4660-MHz (c) and at 4750-MHz (d); and the
FWHM of the spectrum at each point at 4660-MHz (e) and at 4750-MHz (f).
For the identities of the crosses, see Fig.~\ref{fig:fig1}(a).}
\label{fig:fig3}
\end{figure*}

\begin{figure*}
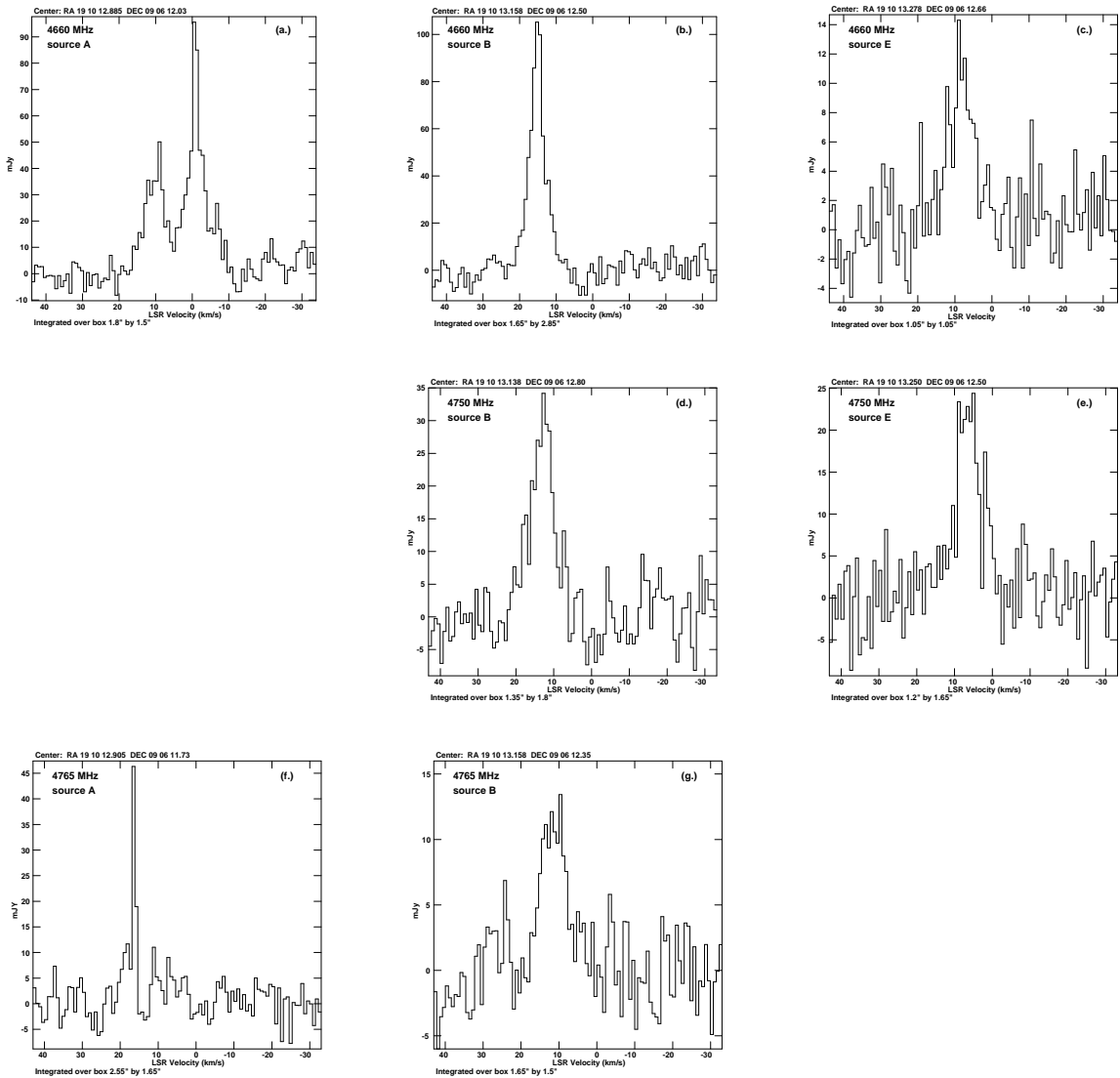

%\centering
\begin{minipage}[c]{.25\textwidth}
   \centering
   \includegraphics[width=\textwidth]{fig4a.eps}
\end{minipage}
\hspace{0.05\textwidth}
\begin{minipage}[c]{.25\textwidth}
   \centering
   \includegraphics[width=\textwidth]{fig4b.eps}
\end{minipage}
\hspace{0.05\textwidth}
\begin{minipage}[c]{.25\textwidth}
   \centering
   \includegraphics[width=\textwidth]{fig4c.eps}
\end{minipage}\\[4truemm]
% want to push d and e over to line up with 4660 
\begin{minipage}[c]{.25\textwidth}
%   \centering
\end{minipage}
\hspace{0.30\textwidth}
\begin{minipage}[c]{.25\textwidth}
   \centering
   \includegraphics[width=\textwidth]{fig4d.eps}
\end{minipage}
\hspace{0.05\textwidth}
\begin{minipage}[c]{.25\textwidth}
%   \centering
   \includegraphics[width=\textwidth]{fig4e.eps}
\end{minipage}\\[4truemm]
\hspace{-0.30\textwidth}
\begin{minipage}[c]{.25\textwidth}
%   \centering
   \includegraphics[width=\textwidth]{fig4f.eps}
\end{minipage}
\hspace{0.05\textwidth}
\begin{minipage}[c]{.25\textwidth}
%   \centering
   \includegraphics[width=\textwidth]{fig4g.eps}
\end{minipage}
\begin{minipage}[c]{.25\textwidth}
%   \centering
\end{minipage}

\caption{Spectra integrated over DMG components.  (a) Source A at 4660-MHz; 
(b) source B at 4660-MHz; (c) Source E at 4660-MHz; (d) source B at 4750
MHz; (e) source E at 4750-MHz; (f) source A at 4765-MHz; (g) source B at
4765-MHz.}
\label{fig:fig4}
\end{figure*}

\begin{figure*}
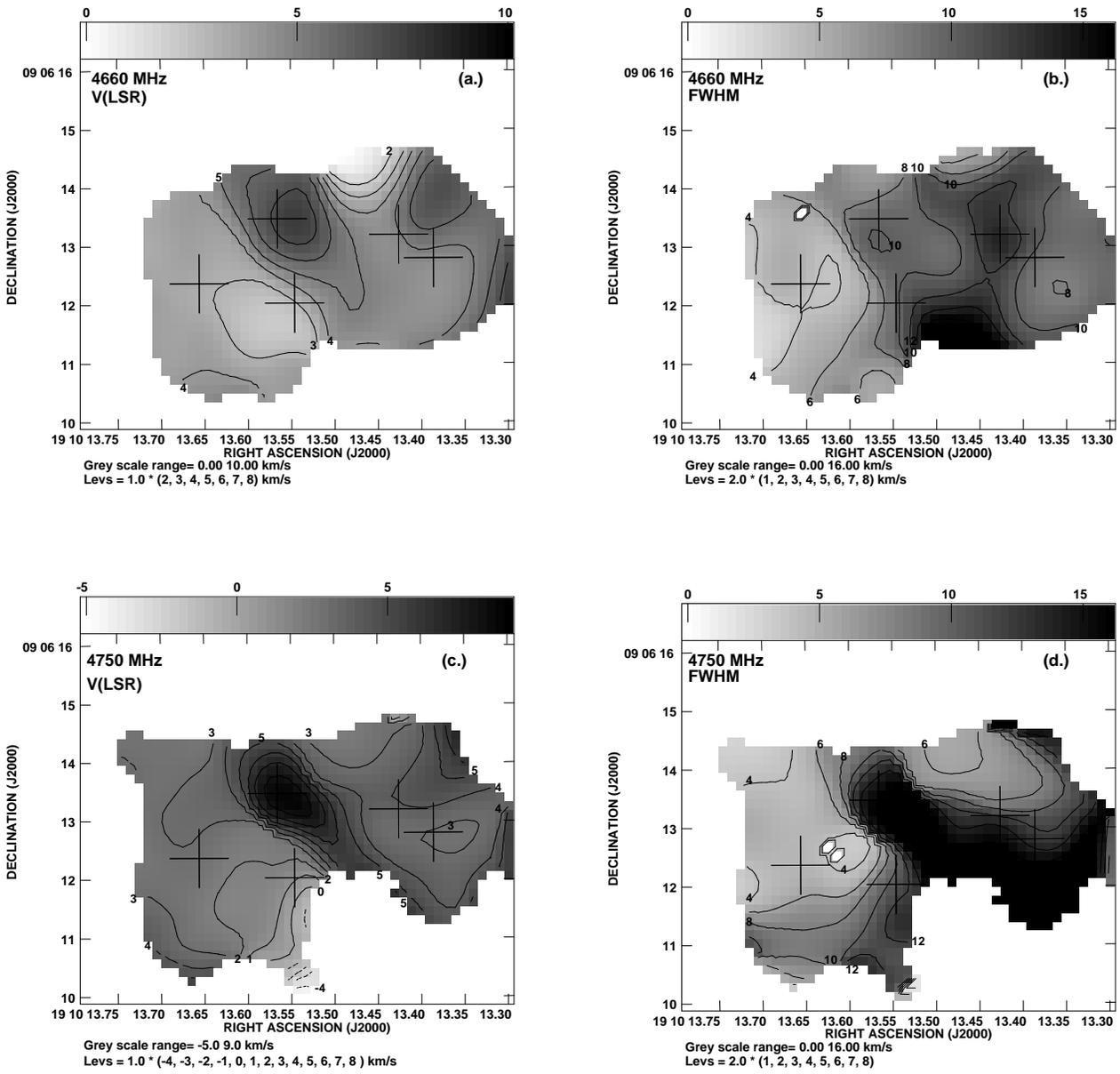

\centering
\begin{minipage}[c]{.45\textwidth}
   \centering
   \includegraphics[width=\textwidth]{fig5a.eps}
\end{minipage}
\hspace{0.05\textwidth}
\begin{minipage}[c]{.45\textwidth}
   \centering
   \includegraphics[width=\textwidth]{fig5b.eps}
\end{minipage}\\[4truemm]
\begin{minipage}[c]{.45\textwidth}
   \centering
   \includegraphics[width=\textwidth]{fig5c.eps}
\end{minipage}
\hspace{0.05\textwidth}
\begin{minipage}[c]{.45\textwidth}
   \centering
   \includegraphics[width=\textwidth]{fig5d.eps}
\end{minipage}
\caption{The \vlsr \ and FWHM variation within source G.  (These panels are
blowups of parts of Fig.~\ref{fig:fig3} with contours superimposed.)  (a)
\vlsr \ at 4660-MHz; (b) FWHM at 4660-MHz; (c) \vlsr \ at 4750-MHz; (d) FWHM at 4750-MHz.}
\label{fig:fig5}
\end{figure*}

\begin{figure*}
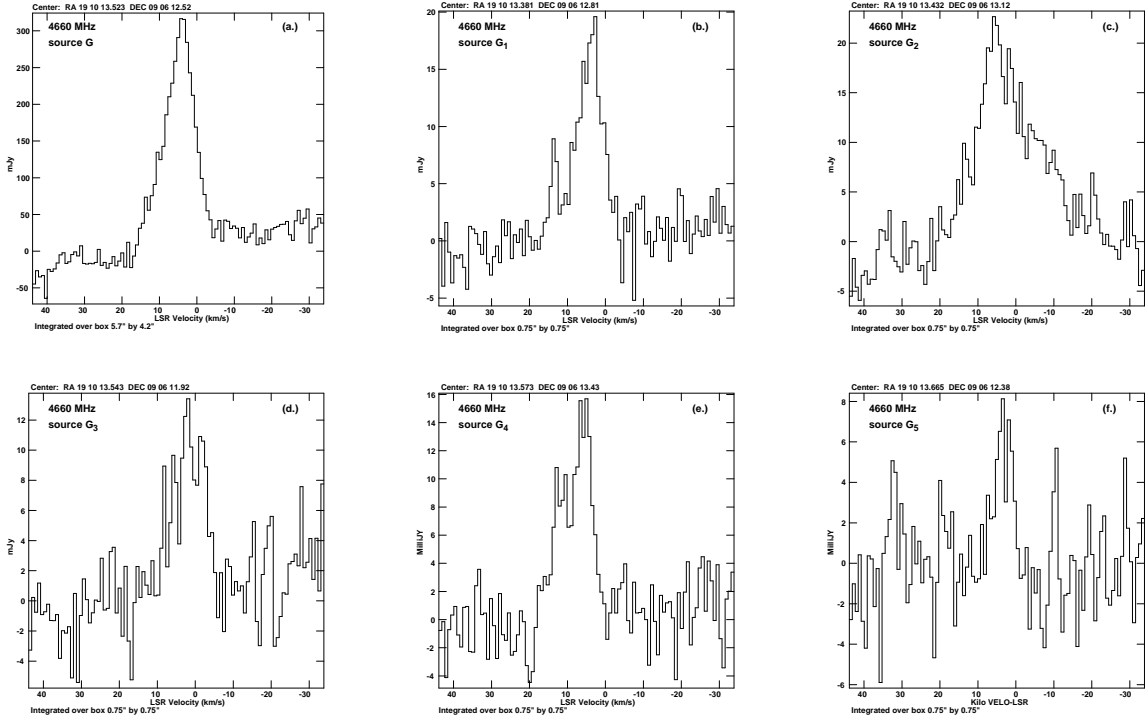

%\centering
\begin{minipage}[c]{.25\textwidth}
   \centering
   \includegraphics[width=\textwidth]{fig6a.eps}
\end{minipage}
\hspace{0.05\textwidth}
\begin{minipage}[c]{.25\textwidth}
   \centering
   \includegraphics[width=\textwidth]{fig6b.eps}
\end{minipage}
\hspace{0.05\textwidth}
\begin{minipage}[c]{.25\textwidth}
   \centering
   \includegraphics[width=\textwidth]{fig6c.eps}
\end{minipage}\\[4truemm]
\begin{minipage}[c]{.25\textwidth}
%   \centering
   \includegraphics[width=\textwidth]{fig6d.eps}
\end{minipage}
\hspace{0.05\textwidth}
\begin{minipage}[c]{.25\textwidth}
%   \centering
   \includegraphics[width=\textwidth]{fig6e.eps}
\end{minipage}
\hspace{0.05\textwidth}
\begin{minipage}[c]{.25\textwidth}
%   \centering
   \includegraphics[width=\textwidth]{fig6f.eps}
\end{minipage}
\caption{Spectra at 4660-MHz integrated over selected regions.  (a)
Integrated over the entire of Source G; (b) -- (f) integrated over 0.75 x
0.75 arcsec$^2$ boxes at the positions of DMG components G$_1$ -- G$_5$.  Several 
spectra are distinctly non-Gaussian and suggest multiplicity.  The spectrum
of G$_2$ clearly contains multiple components.}
\label{fig:fig6}
\end{figure*}

Fig.~\ref{fig:fig1}(d) shows the integrated 4765-MHz emission from the
extended OH region.  The distribution of 4765-MHz emission is similar to
the 4660-MHz and 4750-MHz distributions [Fig.~\ref{fig:fig1}(b) and (c)].
The 4765-MHz image has poorer angular resolution (2 arcsec compared with
1 arcsec).  Also, at 4765-MHz, there are some unresolved velocity components
(FWHM $\leq$1 \kms) which are bright enough to significantly affect the
integrated flux.  In particular, the unresolved velocity feature from the
source G$_4$ region [near \vlsr=2.3 \kms \ in Fig.~\ref{fig:fig2}(c)] is the
cause of the northward bulge of the contours near this position in
Fig.~\ref{fig:fig1}(d).

At 4765-MHz, source A is dominated by a single feature with FWHM$\leq$1
\kms \ [see Fig.~\ref{fig:fig4}(f)].  Emission from source B at this
frequency is less intense (flux density $\sim$10 mJy/beam) than the
emission at 4660-MHz and 4750-MHz [compare Fig.~\ref{fig:fig4}(b), (d), and
(g)].  The position listed in Table~\ref{tab:tab4} and an angular size of 
$\sim$2 arcsec were determined from Gaussian fits.  
Although the values of \vlsr \ are slightly different, all three lines from
source B have comparable linewidths (see Table~\ref{tab:tab4}).

An analysis of source G at 4765-MHz is similar to the analysis for 4660-MHz
and 4750-MHz; Gaussian fitting was not possible due to the inferior angular
resolution and unresolved (in velocity) bright components.  However, it is
apparent that the distribution is complex, and that there are both
point-like components and components that are resolved both spatially and
in velocity.  The spectrum for the whole source G region is shown in
Fig.~\ref{fig:fig8}(a).  Spectra from near the directions of components
G$_1$ and G$_4$ exhibit narrow features (FWHM$\leq$ 1 \kms); see
Fig.~\ref{fig:fig8}(b) and (c).  Parameter values for the sources near G$_1$
and G$_4$ are presented in Table~\ref{tab:tab4}.  Only for G$_4$ do
Gaussian fits yield an non-zero angular size:  $\sim$0.3 arcsec.

To determine the line parameters for the extended emission in source G, a
Gaussian representing the narrow 2.3 \kms \ feature apparent in
Fig.~\ref{fig:fig8}(a) was initially subtracted.  The peak, \vlsr, and FWHM
of the residual are presented in Table~\ref{tab:tab5}.

\subsection{Other 6-cm compact OH sources in the W49A region}

In addition to the sources discussed above in the 15 x 6 arcsec$^2$
region near source G, a number of other OH sources in W49A were detected.
The positions, peak flux densities, \vlsr's, and FWHM's are summarised in
Table~\ref{tab:tab4}.  In most cases, the absolute position error is
dominated by calibration uncertainties ($\sim$0.1\arcsec).  Most of the
sources in Table~\ref{tab:tab4} are unresolved with upper limits in the
range 0.4 -- 1.0 arcsec, depending on the strength of the OH
emission.  Exceptions are two sources discussed above (B and G$_4$) and
South 1.  

South 1 is a rather unusual 4660-MHz maser. With the velocity resolution
used by PGW, this source appeared to be single with angular size $<$1.2
arcsec.  At the epoch of this study, and with the present 0.78 \kms \
resolution, two components with velocities differing by 4.7 \kms \ and
separated by 1.0 arcsec are observed.  The component with \vlsr=13.6 \kms \
appears not to have changed, but the component with \vlsr=18.3 \kms \ was
not present in 1984.  Both components appear to be spatially resolved in
the current study.  A critical question we cannot address with the
currently available data is whether, if observed with better velocity
resolution, each will turn out to be a cluster of point-like components,
similar to 4765-MHz masers in W3(OH) and DR21EX (see PGD).

In all three lines, there is emission within a few arcseconds of J$_1$,
J$_2$, and the bright arc associated with the extended source L [see
Fig.~\ref{fig:fig1}(a) -- (c)].  Emission in the individual lines is not
coincident; and, except for 4660-MHz, the emission is displaced by 
$\sim$1 -- 3 arcsec from the continuum sources.

\section{DISCUSSION}

The broad OH lines (FWHM $\geq$ 5 \kms) have been detected in the
directions of sources B and G in all three OH lines (see
Fig.~\ref{fig:fig1}).  Source G is the primary spatially extended source
with a linear extent of $\sim$0.3 pc ($\sim$5 arcsec).  Below, we first
establish the line of sight location of the OH emitting gas and estimate
the physical conditions in the molecular gas.  Then, we address the
brightness temperatures of the broad lines and show that they are in fact
masers; thus population inversions must occur over a $\sim$0.3 pc region in
contrast to the narrow 4765-MHz lines studied by PGD which are confined to
AU-scale regions.  Finally, we find a striking relationship between the
broad 6-cm OH lines and the 18-cm OH emission lines.

\subsection{Non-maser molecular line and recombination line studies}

Many molecular line studies of W49A have been carried out in the past 30
years.  Typically, molecules in this region emit or absorb in two velocity
ranges: $\sim$4 \kms \ and $\sim$14 \kms.  However, most of the previous
studies do not have sufficient angular resolution to investigate in detail
the source A, B, and G region of interest (15 x 6 arcsec$^2$).  The high
resolution studies of H$_2$CO \citep{dg90}, CS \citep*{sgs93,dwu99}, and
CH$_3$CN \citep{wdwg01} are valuable for determining the line of sight
location of the OH emitting gas with respect to the continuum sources and
for estimates of molecular density, projected density, and neutral gas
kinetic temperature.

\cite{dg90} observed 2- and 6-cm H$_2$CO absorption in W49A with
$\sim$1.5-arcsec resolution.  Toward source G, H$_2$CO absorption
occurred in three broad, blended features.  While H$_2$CO can absorb the
cosmic microwave background, the absorption is so deep that the H$_2$CO
must lie in front of source G.  The parameters of the 4.4 \kms \ H$_2$CO
feature are similar to those of the 6-cm OH emission (see
Table~\ref{tab:tab5}).  Therefore, the molecular gas in the velocity range
in which OH emits lies in front of the HII regions.  Toward sources A and
B, H$_2$CO absorption is observed in more restricted velocity ranges.
Again, the velocity ranges and line widths agree well with those of the
broad OH emission lines.
\cite{dg90} derived a density of $n_{H_2}$=4 10$^5$ cm$^{-3}$ (for a
temperature 40 K).  These authors point out that the 18-cm OH maser
velocities are in the range 2 -- 8 \kms.  Thus, the ground state maser
velocities fall within the velocity extent of the 6-cm OH emission profiles
toward source G [see Fig.~\ref{fig:fig6}(a), Fig.~\ref{fig:fig7}(a),
Fig.~\ref{fig:fig8}(a)].

\cite{sgs93} carried out a multi-transition study of CS and C$^{34}$S
(J=3$\rightarrow$2, 5$\rightarrow$4, and 7$\rightarrow$6) in a
$\sim$45 x 24 arcsec$^2$ region centred on source G.  The angular
resolution ranged from 12 to 20 arcsec, depending on the transition.
Three clumps were identified.  The central and the southwest clumps (CS-C
and CS-SW) were centred near source G and sources A/B, respectively.  For
CS-C, \vlsr \ and the FWHM were similar to the corresponding values for
6-cm OH lines integrated over source G listed in Table~\ref{tab:tab5}; for
CS-SW, \vlsr \ and the FWHM were similar to the corresponding values for
6-cm OH emission from regions A and B listed in Table~\ref{tab:tab4}.  For
an assumed $T_k$=50 K, \cite{sgs93} find $n_{H_2}$ $\sim$10$^6$ cm$^{-3}$
and $N_{H_2}$ $\sim$ 10$^{24}$cm$^{-2}$.

\cite{dwu99} report a study of the CS J=2$\rightarrow$1 line in W49A
with the BIMA array with angular resolution $\sim$4 arcsec.  This study has
better angular resolution than the \cite{sgs93} study; however, because it
involves a lower-lying transition, interpretation is complicated by
possible self-absorption.  The CS J=2$\rightarrow$1 line toward source G
shows absorption at \vlsr $\sim$15 \kms \ and emission at $\sim$4 \kms.
The brightness temperature of the background HII region is 17 K at 6-cm.
Because the 4 \kms \ gas is in front of the HII region (see above), we
interpret this observation to show that the kinetic temperature of the gas
emitting the 6-cm OH lines toward source G is significantly greater than 17 
K.
\cite{dwu99} also report observations of the C$^{34}$S J=3$\rightarrow$2 line with the IRAM telescope.  They observed that toward
source G the
\vlsr=4 \kms \ component was about 50 per cent of the intensity the 12 \kms \
component, while \cite{sgs93} observed these components of
the corresponding J=5$\rightarrow$4 line to have comparable intensity.
\cite{dwu99} suggested that the differing intensity ratios is evidence for 
opacity effects in C$^{34}$S; however, it might also be further evidence
for a higher temperature in the 4 \kms cloud.

\cite{wdwg01} observed the J=12$\rightarrow$11 lines of CH$_3$CN with the BIMA array 
with angular resolution $\sim$0.8 arcsec.  These lines require kinetic
temperatures $\geq$50 K.  These highly excited lines are found in six
discrete regions, including one displaced by 0.6 arcsec from G$_1$ (\vlsr=
0.9$\pm$2 \kms) and another $\sim$0.9 arcsec south of G$_2$ (\vlsr=
3.7$\pm$2
\kms), while extended emission at a lower level is reported around the B and G
regions.  

In summary, the high resolution observations of H$_2$CO, CS, and CH$_3$CN
show that the kinetic temperature of the molecular gas associated with
source G is $\geq$50 K, that the density of H$_2$ is
$\sim$10$^6$ cm$^{-3}$, and that the column density of H$_2$ is
$\sim$10$^{24}$ cm$^{-2}$.  These values are comparable to those
derived from the observations of the 6-cm OH lines (see Section 4.2 and PGD).

Table~\ref{tab:tab6} is a comparison of the positions and velocities of the
the excited OH sources with those of the DMG HII regions.  The separation
between OH sources and DMG continuum components ($\Delta \theta$) are
listed.  The positional comparisons in Table~\ref{tab:tab6} quantify the
visual impression from the moment images in Fig.~\ref{fig:fig1}(b) -- (d): OH
emission in all three lines closely, but imperfectly, follows the continuum
emission.  In most cases, the displacements of the OH line from the
continuum positions are less than the sizes of the continuum components.
The three exceptions are: 1) 4750-MHz near source J$_2$ which as noted
above may be associated with the more diffuse source L; 2) South 1, in
which the OH position falls on extended emission from the HII complex [see
Fig. 1(b) of DMG]; and, 3) source R in which the OH position is displaced
from all continuum components [see Fig. 1(f) of DMG].

It is striking that velocities of the 6-cm OH lines do not agree with those
of the HII regions.  For source G, the OH line velocities are less
positive that those of the ionized gas.  If the OH is on the near side of
the HII complex (as argued above and as expected if the radiation being
amplified arises from the HII regions), the molecular gas is flowing
away from the HII region toward the observer.

\subsection{Brightness temperatures}
To determine a lower limit for the brightness temperatures of the 6-cm OH
lines, the flux density maxima within source G region were located in the
full resolution images.  The maximum at 4660-MHz is $\sim$0.4 arcsec east
of G$_2$; that at 4750-MHz is $\sim$0.7 arcsec south of G$_5$.
Table~\ref{tab:tab7} provides the maximum flux densities and the line
brightness temperatures.  With a larger synthesised beam, PGW determined
lower limits of 1200 K at 4660-MHz and 700 K at 4750-MHz.  Were the lines
spatially unresolved in both studies, the brightness temperature lower
limits would have increased by the ratio of the beam areas, a factor of
$\sim$40.  Because the increase was only a factor $\sim$5, we conclude that
the broad 4660- and 4750-MHz lines from the source G region are
spatially resolved in the current study. Therefore, the values in
Table~\ref{tab:tab7} are estimates of the brightness temperatures and not
lower limits.

At 4765-MHz, the maximum of the broad component cannot be precisely
identified because of the larger synthesised beam and the presence of an
intense narrow line near source G$_2$.  Also we have no earlier VLA data
with which to make comparisons.  In regions displaced from the narrow
components, the lower limit for the brightness temperature exceeds 700 K
(see Table~\ref{tab:tab7}).

The line brightness temperatures in all three lines are so large that we
assume that the populations are inverted and that these are in fact maser
lines.

Based on the studies discussed in Section 4.1, and based on the close
resemblance of the OH emission distributions with the continuum emission
(see Section 3.1), we assume that the OH molecules amplify the continuum
emission from the HII regions.  For the 4660- and 4750-MHz lines we can
make detailed estimates of peak optical depths in the lines and projected
densities of OH and H$_2$.  The continuum brightness temperatures
(extrapolated from the 8-GHz data of DMG) at the corresponding points are
provided in the fourth column of Table~\ref{tab:tab7}.  The fifth column
provides the OH line peak optical depths.  These values are calculated
assuming that the excitation temperatures are small compared to the
continuum brightness.  Although these are maser lines, the gain is small,
unlike that in the narrow 4765-MHz lines (see for example, PGD where gains
corresponding to $\tau_0 \sim$ 20 are observed).

To estimate the projected density of OH required to provide this gain, we
follow PGD\footnotemark[3]\footnotetext[3]{Note that equation (1) of PGD is 
incorrect.  It should read: \[ \tau_0 = \sqrt{\frac{4 ln(2)}{\pi}} \frac{c^2}{8 \pi
\nu^2}\frac{g_u A_{ul}}{\Delta \nu}\Delta n L. \]  Therefore the
coefficients of equations(3) and (5)---(7) should be increased by a factor
of 9.  Finally, the un-numbered equation following from equation(7) should
read\[ L_{AU} = \frac{2 \ 10^8}{n_{H_2}}.  \] The required density in the
next sentence should read n$_{H_2} \geq 2 \ 10^5$
\ cm$^{-3}$.}.   For the broad lines, we use the
linewidth 4 \kms \ (the smallest values in the region).  We estimate that
the inversion efficiency lies in the range 0.01 -- 0.1, and that the fraction
of OH molecules in the $^2\Pi_{1/2}, J=1/2$ state is in the range 0.03 --
0.1 (corresponding to rotational excitation temperatures 50 -- 200 K).
For the peak optical depths of the 4660- and 4765-MHz lines, we obtain:
\begin{equation}
\tau_0 = 3.5 \ 10^{-17} N_{OH};
\end{equation}
and, for the 4750-MHz line:
\begin{equation}
\tau_0 = 7.0 \ 10^{-17} N_{OH}.
\end{equation}

The abundance of OH relative to H$_2$ is uncertain.  A number of
determinations for relatively diffuse clouds obtain 
$\frac{n_{OH}}{n_{H_2}}$ $\sim$10$^{-8}$ (e.g. \citealt{c79};
\citealt{ll96}).  For more opaque clouds, values a factor of 10
larger are obtained (e.g. \citealt{jfgw94}; \citealt*{hww00}).  For strong
masers, values 10 -- 100 times larger are usually invoked
(e.g. \citealt{pk96} and references therein).  Because the clouds from
which we observe the broad OH lines most resemble the opaque clouds, we
adopt a value of $\frac{n_{OH}}{n_{H_2}}$= 2 $\times$ 10$^{-7}$.  Finally,
we write equations (1) and (2) in terms of $N_{H_2}$.  For the 4660- and
4765-MHz lines:
\begin{equation}
\tau_0 = 7.0 \ 10^{-25} N_{H_2}, 
\end{equation}
and for the 4750-MHz line:
\begin{equation}
\tau_0 = 1.4 \ 10^{-24} N_{H_2}. 
\end{equation}

The column densities of H$_2$ required to account for the gains in the
fifth column of Table~\ref{tab:tab7} are provided in the final column.  These
column densities are similar to those derived in the CS study by
\cite{sgs93}.

Finally, we investigate the consequences of assuming that the OH is on the
far side of the HII regions.  The background temperature would be $\sim$10
K, leading to optical depths of $\sim$6.2 which would require an increase
of OH column density and therefore H$_2$ column density by a factor
$\sim$15.  Unfortunately such an increased column density cannot be ruled
out because of the uncertainty in $\frac{n_{OH}}{n_{H_2}}$.  Therefore our
conclusion that the OH observed is on the near side of the HII regions
rests on 1) the H$_2$CO absorption data, and 2) on the similar shapes of
the OH emission distribution and the continuum emission.

\subsection{Relationship with 18-cm OH masers}

Several authors have studied maser lines from ground state (18-cm) OH and
from H$_2$O in W49A with milli-arcsecond resolution.  Many OH masers in all
four ground state lines have been discovered near source G
(e.g. \citealt{gm87}).  A five epoch proper motion study of H$_2$O masers
established that the dynamical centre of an exceptionally powerful outflow
is located in source G \citep{gmr92,dwgwm00,wdwg01}.  The resolutions of
these 18-cm OH and H$_2$O studies correspond to AU linear scales which are
too small to use for determining physical conditions in the extended
molecular gas ($\sim$0.1 pc scales).

\cite{gm87} catalogued 367 18-cm OH masers in this region.  
Fig.~\ref{fig:fig10}(a) shows the locations of the masers spots in all four
transitions superimposed on the 4660-MHz OH first moment image and the
continuum distribution.  There are prominent similarities and differences
between the distribution of 4660-MHz emission and the 18-cm maser spots.
For sources A and B, the 18-cm OH and the 4660-MHz emission are similar.
However, for source G, the 18-cm maser spots tend to follow the outer edge
of the 4660-MHz distribution (and of the continuum distribution).  The
phenomenon that 18-cm OH masers associated with compact HII regions tend to
fall on the edges of the HII regions is well known (see e.g.,
\citealt{gm87}).  One possible explanation is that the masers are located in
molecular gas which is located on the far side of the HII region.  Because
compact HII regions are optically thick at 18-cm, such masers would be
attenuated and not detected.  In this view, the concentration at the edges
is only apparent due to obscuration of others behind the HII region.
However, if this were the correct explanation for the difference in
distributions, we would expect that in about 50 per cent of the cases the
molecular gas containing the masers would be on the near side of the HII
region.  Such masers would appear to be distributed across the face of the
HII region, and these cases are uncommon.  The second explanation is based
an expanding spherical HII region.  At the edges, lines of sight passing
through the sphere have maximum path-lengths in which OH has similar
physical conditions and in which the velocity coherence is most favourable.
An objection to this explanation is that turbulence in a region of
massive star formation may be so large that an organized, spherical outflow
is not realized.

Fig.~\ref{fig:fig10}(b) is a histogram of the velocities of the 348 18-cm
masers in the direction of source G.  The maser velocities are grouped into
bins 1.5 \kms \ wide, and the number per velocity interval is plotted.
This plot represents the velocity field in the gas emitting the 18-cm lines
independently of the the intensity of the individual features.  A simple
count of masers unweighted by intensity is of most interest because the
high gain 18-cm masers depend on physical conditions at the size scale of
the maser spots; such small regions will be unimportant in the
formation of the spatially extended broad 6-cm lines.  The shape of this
distribution is similar to the line shapes of the broad 6-cm lines [compare
with Fig.~\ref{fig:fig2}(a) -- (c)].  Thus, the velocity distribution of the
gas emitting the 18-cm masers is similar to that of the gas emitting the
broad 6-cm lines as well as the CS lines (see Fig. 7 of
\citealt{sgs93}).  The ionized gas and the molecular gas do not have
the same average velocity (see Section 4.2) and the gas emitting 18-cm
lines and that emitting the 4660-MHz line are in separate regions [as shown
by Fig.~\ref{fig:fig10}(a)].  Therefore, the similarity of the velocity
distributions is surprising, unless all of molecular gas is part of an
organized outflow.

\subsection{Conclusions}

We have presented observations of the broad 6-cm OH lines.  These lines are
maser lines with linear extents and velocity distributions comparable to
the those of several non-maser molecules in this region.  However, the low
gain, broad 6-cm masers have the same velocity distribution as the high
gain 18-cm masers.  This similarity suggests that all of the observed OH
masers are part of the same outflow from source G, although the broad 6-cm
and the 18-cm masers must be located in distinct regions of the outflow.

Several outstanding problems remain.  First, an explanation must be found
for the excitation of the broad, low gain masers throughout the $\sim$0.3
pc diameter region.  Second, an explanation must be offered for why we
observe that only a small fraction of the total area from which the broad
lines arise contains spots with large gains in the 4765-MHz and in the
18-cm lines.  In order to resolve the second problem, we suggest that long
lines of sight through the turbulent molecular gas may be likely along
which the velocity is coherent to within an observed linewidth.  New models
of turbulent gas in regions of massive star formation are required which
provide estimates of the number of such lines of sight that will be
expected in a given area at one time, and further provide estimates of how
long these coherent lines of sight will exist.

\section{ACKNOWLEDGMENTS}
P. Palmer thanks the NRAO for hospitality during 
several extensive visits while this work was carried out.

\begin{table*}
\begin{minipage}{125mm}
\caption{Pointing positions and central \vlsr.}
\label{tab:tab1}
\begin{tabular}{lccccc} \hline
Year & \multicolumn{4}{c}{Pointing Position} & Central \vlsr \\
     & \multicolumn{2}{c}{B1950.0} & \multicolumn{2}{c}{J2000.0} & (\kms)
\\ \hline
2001 & $19^{\rmn{h}} 07^{\rmn{m}} 47\fs319$ & $09\degr 00\arcmin 20\farcs53$ &
$19^{\rmn{h}} 10^{\rmn{m}} 10\fs948$ & $09\degr 05\arcmin 17\farcs67$ & +8.23 \\
2002 & $19^{\rmn{h}} 07^{\rmn{m}} 49\fs500$ & $09\degr 01\arcmin 15\farcs00$ &
$19^{\rmn{h}} 10^{\rmn{m}} 13\fs112$ & $09\degr 06\arcmin 12\farcs29$ & +5.0 \\ \hline
\end{tabular}
\end{minipage}
\end{table*}

\begin{table*}
\begin{minipage}{125mm}
\caption{Observing parameters.}
\label{tab:tab2}
\begin{tabular}{lccccccc}
\hline 
Date & Line & Array &  $\Delta$V  & Bandpass &Time on & RMS & Synth. Beam \\
 & (MHz) &   & (km s$^{-1}$) & Calibrator & Source & (mJy/bm) &(arcsec x 
arcsec, PA(\degr))
\\ \hline
2001 Aug 27 & 4660 & C &  0.785 & 3C286 & 38 min &  3.5 & 6.6 x 4.4, -53 \\
2001 Nov 15 & 4750 & D &  0.770 & --- & 35 min &  3.9 & 19. x 14., 56 \\
2002 Apr 01 & 4660 & A &  0.785 & 3C286 & 169 min & 1.4 & 0.44 x 0.42, -90 \\ 
2002 Apr 01 & 4750 & A &  0.770 & 3C286 & 166 min &  1.3 & 0.43 x 0.52, 76 \\ 
2002 Jun 01 & 4765 & BnA & 0.768 & 3C84 & 78 min &  2.1 &1.8 x 0.52, 60 \\ \hline
\end{tabular}
\end{minipage}
\end{table*}

\begin{table*}
\begin{minipage}{125mm}
\caption{Peak observed flux densities of the broad 6-cm OH lines in W49A.
Estimated uncertainties are listed.}
\label{tab:tab3}
\begin{tabular}{ccccccc}
\hline 
Observers & Date & Instrument & Beamwidth &  4660-MHz &  4750-MHz & 4765
MHz \\
 & & & (arcsec)  & (mJy)  & (mJy) & (mJy)\\
\hline
RPZ & 1974 February & Effelsberg & 156 & 300 & 350 & 200 \\
GM & 1982 February/May & Effelsberg & $\sim$180 & 380(20) &340(20) &
0.20(2) \\
PGW & 1984 March & VLA & 25.5 x12 & 400(20) & 320(20) & ---\\
CMW & 1989 July & Lovell & 216 & 400 & 300 & 200 \\
Smits & 1994 October/November & Harebeesthoek & 600 & $\leq$1800 & $\leq$900 &
--- \\
(this paper) & 2001 August 27 &VLA &  22.8 x15.6 & 380(40) & ---
 & --- \\
(this paper) & 2001 November 15 & VLA &  40.5 x31.5 & --- & 410(40) &---\\
(this paper)& 2002 April 1 &  VLA & 15 x6 & 350(40) & 300(30) & --- \\ 
(this paper)& 2002 June 1 &  VLA & 15 x6 &---  & ---  & 190(30) \\ \hline
\end{tabular}
\end{minipage}
\end{table*}

\begin{table*}
\caption{Parameters for compact OH Sources.  The 1-$\sigma$ uncertainties
are listed for the peak flux density, \vlsr, and FWHM.  For estimated
uncertainties in positions, see text.}
\label{tab:tab4}
\begin{minipage}{125mm}
\begin{tabular}{ccccccc}
\hline
 Nearest &Line&  $\alpha$(J2000.0) & $\delta$(J2000.0) &  Peak & V$_{LSR}$ & FWHM  \\
DMG Source &(MHz)  &  & 
&  (mJy/beam) & (\kms) &(\kms)  \\ \hline
A     & 4660 & $19^{\rmn{h}} 10^{\rmn{m}} 12\fs89$ & $09\degr 06\arcmin 11\farcs9$ & 30(3) & 10.1(3) & 7.3(8)  \\
A     & 4660 & $19^{\rmn{h}} 10^{\rmn{m}} 12\fs90$ & $09\degr 06\arcmin 12\farcs0$ & 70(4) &  -0.9(1) & 4.3(3)  \\
A     & 4765 & $19^{\rmn{h}} 10^{\rmn{m}} 12\fs89$ & $09\degr 06\arcmin 12\farcs1$ & 64(3) & 16.3(3) & $\leq$1.0  \\
B     & 4660 & $19^{\rmn{h}} 10^{\rmn{m}} 13\fs17$ & $09\degr 06\arcmin 12\farcs7$ & 55(6) & 15.2(2) & 4.1(5) \\  
B     & 4750 & $19^{\rmn{h}} 10^{\rmn{m}} 13\fs14$ & $09\degr 06\arcmin 12\farcs7$ & 18(4) & 13.1(2) &7.4(3) \\
B     & 4765 & $19^{\rmn{h}} 10^{\rmn{m}} 13\fs16$ & $09\degr 06\arcmin 13\farcs1$ & 10(2) & 12.4(2) & 6.2(5)\\
E     & 4750 & $19^{\rmn{h}} 10^{\rmn{m}} 13\fs25$ & $09\degr 06\arcmin 12\farcs4$ & 15(4) & 6.1(3) & 8.3(3) \\
G$_1$ & 4765 & $19^{\rmn{h}} 10^{\rmn{m}} 13\fs36$ & $09\degr 06\arcmin 12\farcs7$ & 12(3) &-8.0(3)  & $\leq$1.0  \\
G$_4$ & 4765 & $19^{\rmn{h}} 10^{\rmn{m}} 13\fs52$ & $09\degr 06\arcmin 13\farcs9$ & 264(3) & 2.3(3) & $\leq$1.0 \\
H     & 4765 & $19^{\rmn{h}} 10^{\rmn{m}} 13\fs63$ & $09\degr 06\arcmin 17\farcs8$ & 24(3)  & 10.7(4)  & $\leq$1.0 \\
J$_1$ & 4660 & $19^{\rmn{h}} 10^{\rmn{m}} 14\fs14$ & $09\degr 06\arcmin 25\farcs2$ & 16(3) & 7.4(2) & 2.0(10) \\
J$_2$ & 4750 & $19^{\rmn{h}} 10^{\rmn{m}} 14\fs32$ & $09\degr 06\arcmin 22\farcs4$ & 55(2) &4.2(4)  &$\leq$1.0 \\
J$_2$ & 4765 & $19^{\rmn{h}} 10^{\rmn{m}} 14\fs13$ & $09\degr 06\arcmin 27\farcs6$ & 34(2) & 7.5(3) & $\leq$1.0 \\
O     & 4765 & $19^{\rmn{h}} 10^{\rmn{m}} 16\fs36$ & $09\degr 06\arcmin 06\farcs6$ & 16(2) & -1.5(4) & 5.8(10) \\
R     & 4765 & $19^{\rmn{h}} 10^{\rmn{m}} 10\fs94$ & $09\degr 05\arcmin 17\farcs6$ & 77(2) & 11.9(3) &$\leq$1.0 \\ 
South 1&4660 & $19^{\rmn{h}} 10^{\rmn{m}} 21\fs68$ & $09\degr 05\arcmin 01\farcs0$ & 70(2)  & 13.6(3) &  $\leq$1.5 \\  
South 1&4660 & $19^{\rmn{h}} 10^{\rmn{m}} 21\fs68$ & $09\degr 05\arcmin 02\farcs1$ & 51(2) & 18.3(3) & $\leq$1.0 \\ \hline
\end{tabular}
\end{minipage}
\end{table*}

\begin{table}
\caption{Parameters of broad 6-cm OH lines from source G region.  The
1-$\sigma$ uncertainties are listed.}
\label{tab:tab5}
\begin{tabular}{cccc}
\hline \hline
Line & S$_{max}$ & V$_{LSR}$ & FWHM   \\
 (MHz)    & (mJy) & (\kms) & (\kms) \\ \hline
4660  & 310(12) & 3.8(3)  & 9.0(8)   \\
4750  & 270(10) & 3.1(3)  & 6.9(8)  \\
4765  & 140(9)  & 4.2(4) & 10.2(8)  \\
\hline
\end{tabular}
\end{table}

\begin{table*}
\caption{Comparison of V$_{LSR}$: 6-cm OH vs. the ionized gas.}
\label{tab:tab6}
\begin{minipage}{125mm}
\begin{tabular}{cccccccccc}
\hline
Nearest & \multicolumn{2}{c}{4660-MHz} & \multicolumn{2}{c}{4750-MHz} &
\multicolumn{2}{c}{4765-MHz} & H92$\alpha$&  H66$\alpha$& H52$\alpha$ \\
DMG & V$_{LSR}$ & $\Delta \theta$& V$_{LSR}$ & $\Delta \theta$ &
V$_{LSR}$ & $\Delta \theta$ 
&V$_{LSR}$&V$_{LSR}$&V$_{LSR}$ \\
Source & (\kms) & (arcsec) &(\kms) & (arcsec) &(\kms) & (arcsec) & (\kms) & (\kms)& (\kms)\\ \hline
A & 10.1, -0.90& 0.10, 0.18 & -- & -- & 16.3  &0.27 & --& 14.4& 13.1 \\
B & 15.2 &0.10 & 13.1 & 0.10 & 12.4 & 0.30 & 2.5& 19.0& 16.5 \\
E & -- & -- & 6.1 & 0.28  &-- &-- &  9.1& --& -- \\
G$_1$  & -- &-- &-- &-- & -8.0 & 0.12 & 11.8& 7.4& -- \\
G$_4$  & -- &-- &-- &-- &2.3 & 0.47 & 9.7& 10.3& -- \\
G$_{all}$ & 3.8 & ... &3.1  & ... &4.2 & ... & 9.5& 10.4& 10.1 \\
H & -- &-- &-- &-- & 10.7 & 0.64 & 5.2& --& -- \\
J$_1$ & 8.4 & 0.03 &-- &-- &--&--  & --& --& -- \\
J$_2$ & -- & --  & 4.2  &2.70 &7.5  & 0.98 & 7.2& --& --\\
%L &  & & & & & & 1.8,3.2/ 0.7/ -2.3 \\
O  & -- &-- &-- & --&-1.5 &0.49 & 3.8&-1.6 & -- \\
R  &--  &-- &-- &-- &8.6, 11.9 & 3.03 & 6.2& --& -- \\
South 1  & 13.6, 18.3 & 1.14, 1.54 &-- &-- &-- &-- &8.5, 10.5& --& -- \\ \hline
\end{tabular}
\end{minipage}
\end{table*}

\begin{table}
\caption{Brightness temperatures for broad 6-cm OH components}
\label{tab:tab7}
\begin{tabular}{cccccc}
\hline \hline
Line & S$_{max}$ & T$_b$ & T$_c(m)$ &  $\tau_0$ & N$_{H_2}$  \\ 
 (MHz) & (mJy/bm) &  (K)  &  (K)  &  &  (cm$^{-2}$) \\ \hline
4660  & 20  & 5900  &  12800  & -0.38 & 5.4 10$^{23}$ \\
4750   & 15   & 4700  & 7400   & -0.49 & 3.5 10$^{23}$ \\
4765   & $\sim$15  & $\geq$730  &  --   &  --   & --   \\
\hline
\end{tabular}
\end{table}

\centering

\begin{figure*}
%\centering
\begin{minipage}[c]{.25\textwidth}
   \centering
   \includegraphics[width=\textwidth]{fig7a.eps}
\end{minipage}
\hspace{0.05\textwidth}
\begin{minipage}[c]{.25\textwidth}
   \centering
   \includegraphics[width=\textwidth]{fig7b.eps}
\end{minipage}
\hspace{0.05\textwidth}
\begin{minipage}[c]{.25\textwidth}
   \centering
   \includegraphics[width=\textwidth]{fig7c.eps}
\end{minipage}\\[4truemm]
\begin{minipage}[c]{.25\textwidth}
%   \centering
   \includegraphics[width=\textwidth]{fig7d.eps}
\end{minipage}
\hspace{0.05\textwidth}
\begin{minipage}[c]{.25\textwidth}
%   \centering
   \includegraphics[width=\textwidth]{fig7e.eps}
\end{minipage}
\hspace{0.05\textwidth}
\begin{minipage}[c]{.25\textwidth}
%   \centering
   \includegraphics[width=\textwidth]{fig7f.eps}
\end{minipage}
\caption{Spectra at 4750-MHz integrated over selected regions.  (a)
Integrated over the entire of Source G; (b) -- (f) integrated over 0.75 x
0.75 arcsec$^2$ boxes at the positions of DMG components G$_1$ -- G$_5$.
Several spectra are distinctly non-Gaussian and suggest multiplicity.}
\label{fig:fig7}
\end{figure*}

\begin{figure*}
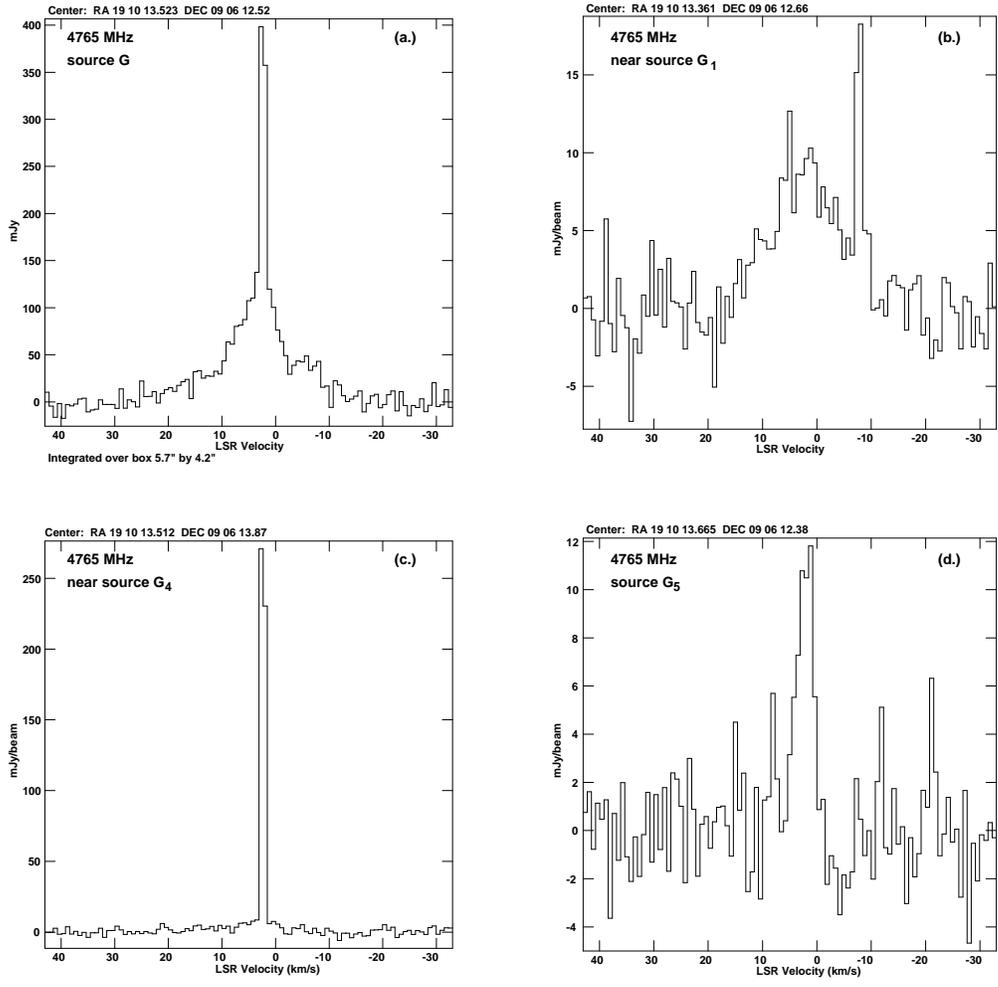

%\centering
\begin{minipage}[c]{.35\textwidth}
   \centering
   \includegraphics[width=\textwidth]{fig8a.eps}
\end{minipage}
\hspace{0.05\textwidth}
\begin{minipage}[c]{.35\textwidth}
   \centering
   \includegraphics[width=\textwidth]{fig8b.eps}
\end{minipage}\\[4truemm]
\begin{minipage}[c]{.35\textwidth}
   \centering
   \includegraphics[width=\textwidth]{fig8c.eps}
\end{minipage}
\hspace{0.05\textwidth}
\begin{minipage}[c]{.35\textwidth}
   \centering
   \includegraphics[width=\textwidth]{fig8d.eps}
\end{minipage}
\caption{Spectra of source G at 4765-MHz.  (a) Integrated over source
G; (b) spectrum at the maximum near component G$_1$ (the narrow component at
\vlsr$\sim$-8 \kms is the point-like component; the broad component is extended);
(c) spectrum at the maximum near component G$_4$; (d) spectrum at the
position of G$_5$.}
\label{fig:fig8}
\end{figure*}

\begin{figure*}
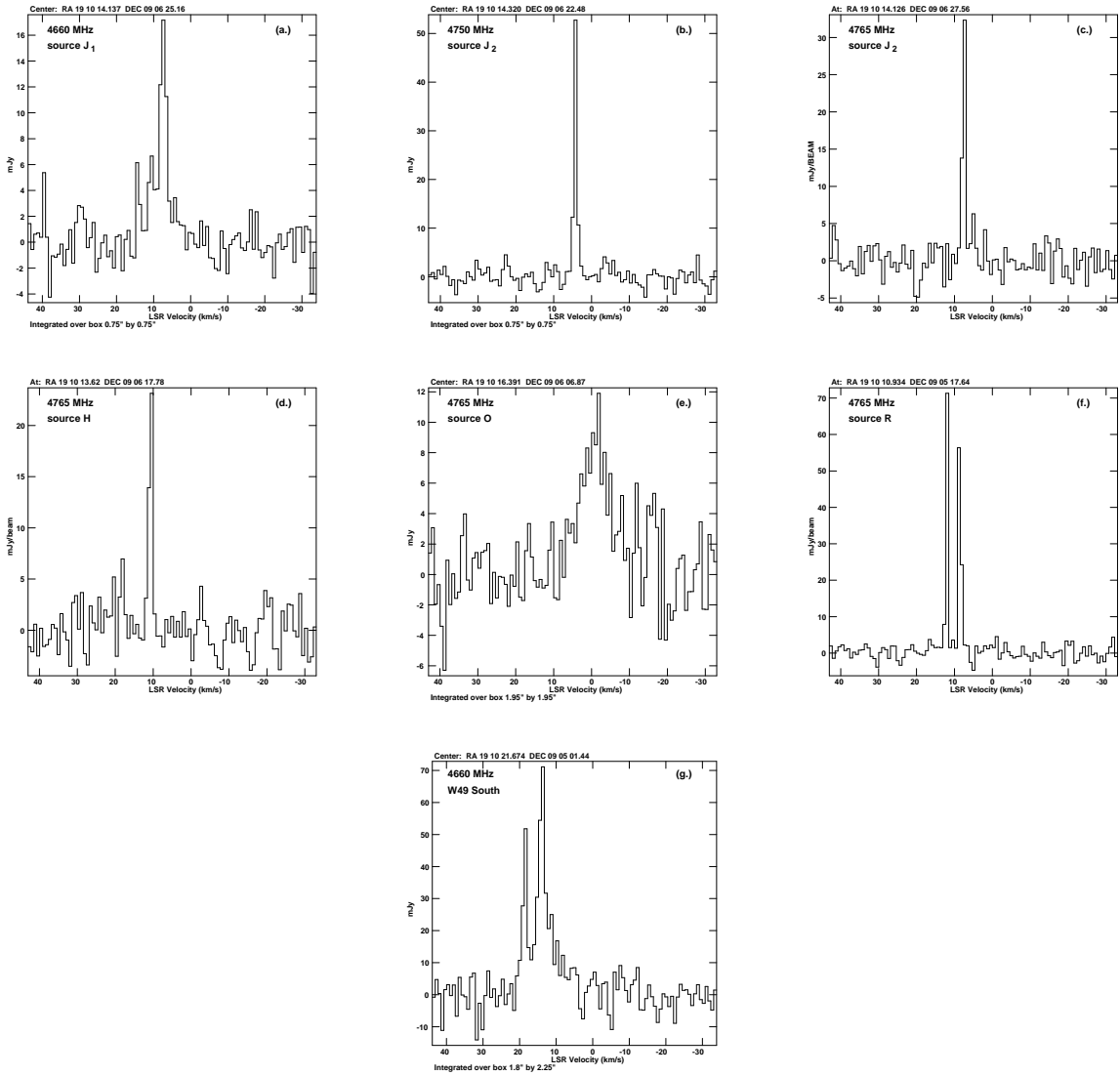

%\centering
\begin{minipage}[c]{.25\textwidth}
   \centering
   \includegraphics[width=\textwidth]{fig9a.eps}
\end{minipage}
\hspace{0.05\textwidth}
\begin{minipage}[c]{.25\textwidth}
   \centering
   \includegraphics[width=\textwidth]{fig9b.eps}
\end{minipage}
\hspace{0.05\textwidth}
\begin{minipage}[c]{.25\textwidth}
   \centering
   \includegraphics[width=\textwidth]{fig9c.eps}
\end{minipage}\\[4truemm]
\begin{minipage}[c]{.25\textwidth}
    \includegraphics[width=\textwidth]{fig9d.eps}
\end{minipage}
\hspace{0.05\textwidth}
\begin{minipage}[c]{.25\textwidth}
%   \centering
   \includegraphics[width=\textwidth]{fig9e.eps}
\end{minipage}
\hspace{0.05\textwidth}
\begin{minipage}[c]{.25\textwidth}
   \centering
   \includegraphics[width=\textwidth]{fig9f.eps}
\end{minipage}\\[4truemm]
\begin{minipage}[c]{.25\textwidth}
   \centering
   \includegraphics[width=\textwidth]{fig9g.eps}
\end{minipage}
\caption{Spectra of sources outside of the extended OH region.  (a) near
source J$_1$ at 4660-MHz; (b) near source J$_2$ at 4750-MHz (this position
could also be considered to be associated with the bright arc of source L);
(c) near source J$_2$ at 4765-MHz.  The positions of the spectra shown in
(a) -- (c) are separated by $\sim$ 2 arcsec.  (d) near source H at
4765-MHz; (e) near source O at 4765-MHz; (f) near R at 4765-MHz; (g)
4660-MHz source near W49 South.}
\label{fig:fig9}
\end{figure*}

\begin{figure*}
\centering
\begin{minipage}[c]{.6\textwidth}
   \centering
   \includegraphics[width=\textwidth,angle=-90]{fig10a.eps}
\end{minipage}\\[4truemm]
\begin{minipage}[c]{.6\textwidth}
   \centering
   \includegraphics[width=\textwidth]{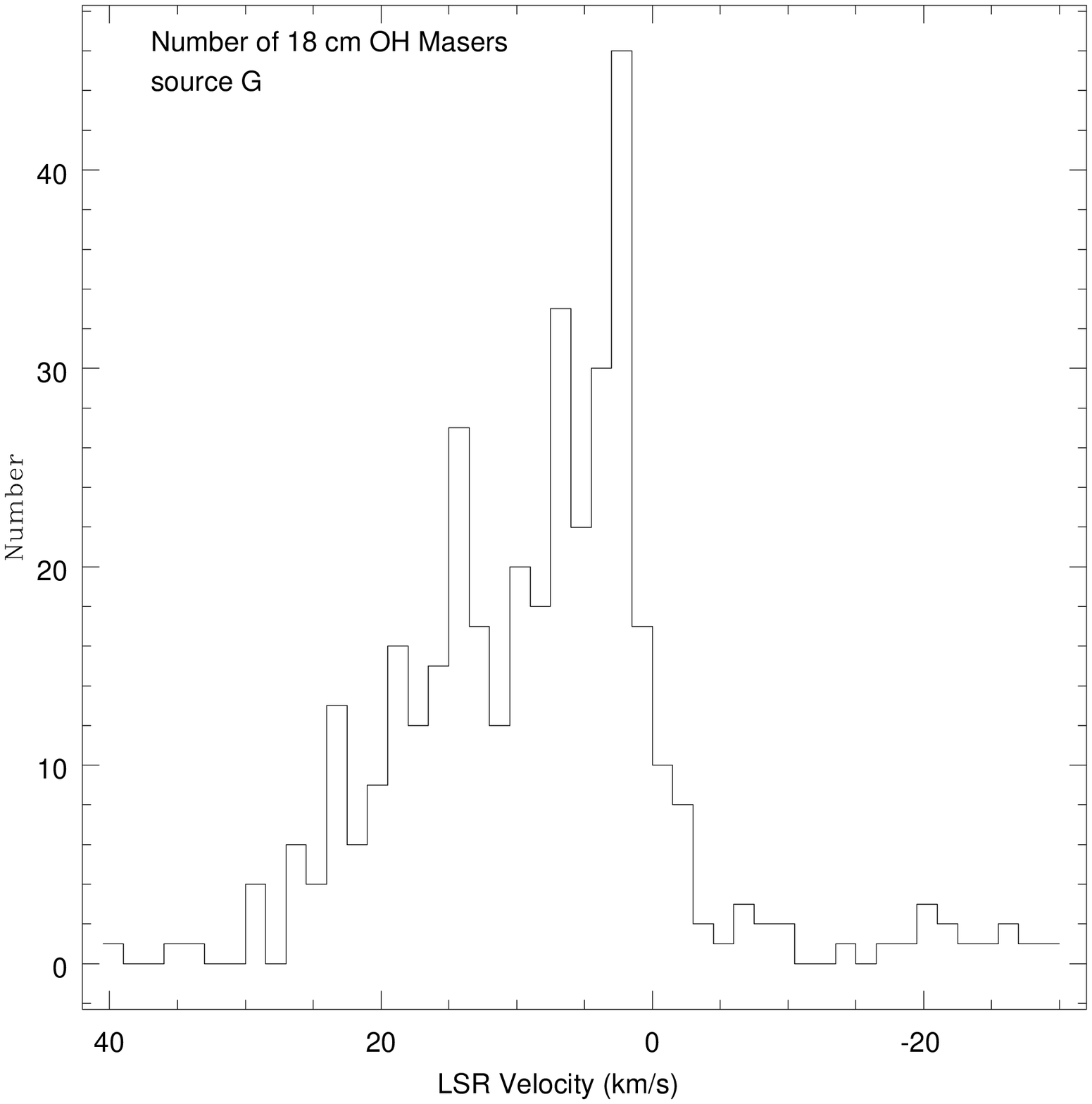}
\end{minipage}
\caption{(a) Distribution of 18-cm masers from Gaume \& Mutel (1987)
superimposed on the 4660-MHz distribution (contours) and on the 3.6-cm continuum
distribution (greyscale) as in Fig.~\ref{fig:fig1}.  The symbols used to identify
masers in the four 18-cm lines are shown on the left side of the figure.
(b) A histogram of the velocity distribution of the 18-cm masers in this
region.  See text for details.}
\label{fig:fig10}
\end{figure*}

\end{document}